# Quasi-monoenergetic Electron Beams from Laser-plasma Acceleration by Ionization-induced Injection in Low-density Pure Nitrogen

Mengze Tao, Nasr A. M. Hafz[*], Song Li, Mohammad Mirzaie, Liming Chen, Fei He, Ya Cheng, and Jie Zhang [¶]

Manuscript received August 3, 2014. This work was supported by the 973 National Basic Research Program of China (Grant No. 2013CBA01504), and the Natural National Science Foundation of China NSFC (Grant Nos: 11121504, 11334013, 11175119, and 11374209).

Mengze Tao and Liming Chen are with the Beijing National Laboratory of Condensed Matter Physics, Institute of Physics, Chinese Academy of Sciences, Beijing 100190, China.

Nasr A. M. Hafz, Song Li, Mohammad Mirzaie and Jie Zhang are with the Key Laboratory for Laser Plasmas (MOE) and Department of Physics and Astronomy, Shanghai Jiao Tong University, Shanghai 200240, China.

F. He and Y. Cheng are with the State Key Laboratory of High Field Laser Physics, Shanghai Institute of Optics and Fine Mechanics, Chinese Academy of Sciences, Shanghai 201800, China.

[*]nasr@sjtu.edu.cn
[¶]jzhang1@sjtu.edu.cn

*Abstract*——**We report a laser wakefield acceleration of electron beams up to 130 MeV from laser-driven 4-mm long nitrogen gas jet. By using a moderate laser intensity ($\sim 3.5 \times 10^{18} W \cdot cm^{-2}$) and relatively low plasma densities ($0.8 \times 10^{18}\ cm^{-3}$ to $2.7 \times 10^{18}\ cm^{-3}$) we have achieved a stable regime for laser propagation and consequently a stable generation of electron beams. We experimentally studied the dependence of the drive laser energy on the laser-plasma channel and electron beam parameters. The quality of the generated electron beams is discussed within the framework of the ionization-induced injection mechanism.**

## I. INTRODUCTION

Laser wakefield acceleration (LWFA) is foreseen as a promising scheme for the development of compact next generation electron beam accelerators. Such table-top laser-based accelerators with acceleration gradients larger than $100 GeV/m$, have a great potential to replace the high-energy traditional accelerators which are huge and expensive [1]. In the last decade, experimental LWFA research has brought significant progress in the enhancement of the electron beam quality, stability, control and maximum energy. A major breakthrough was achieved in 2004 when the first results on quasimonoenergetic beam generation with peak energies up to $170\ MeV$ were demonstrated [1]-[3]. Recent experiments have demonstrated the generation of high quality monoenergetic electron beams with different energies and fluxes, through the interaction of laser pulses having powers higher than 10 TW with gases of hydrogen or helium using gas jets [5]-[6], gas cells [7], and capillary discharge waveguides [8]. Very recent results have shown the generation of electron beams up to $2\ GeV$ of energy using pettawatt laser pulses driving 7 cm-long gas cell of sub-$10^{18}\ cm^{-3}$ density [9].

In the laser wakefiled acceleration scheme (LWFA), plasma wave excitation relies on the ponderomotive force of an ultrashort, intense laser pulse expelling electrons out of the high laser intensity regions. For laser intensities higher than $10^{18}\, W/cm^{-2}$, electrons can be completely expelled out of the intense laser volume and self-trapped in the accelerating fields associated with the plasma wave. Normally, to excite efficient plasma wave for acceleration in the wake of a terawatt (TW) laser having a pulse duration of ~100 $fs$, the plasma should be underdense of density $n_e$ higher than $10^{18} cm^{-3}$. At this density, the electron dephasing length $d_{ph}$ is about 1$cm$ [10] and the electrons could gain energy of about 1 $GeV$ [8] within 1-cm long plasma. To accelerate electrons to multi-$GeV$ energy in a laser plasma accelerator, the acceleration length has to be extended to multi-$cm$, which can be realized by reducing the plasma density to ~ $10^{17}\, cm^{-3}$ and by employing capillary discharges or gas cells. Achieving a high quality electron beam by the LWFA depends on a proper choice of the laser –plasma parameters such as focal spot size, plasma density, interaction length and laser pulse duration etc. For example, a large focal spot leads to longer interaction length and wakefield excitation. A laser pulse that is tens of femtoseconds in length ensures that the laser pulse propagates plasma in a very stable way (i.e. stable against parametric instabilities). Recently, ionization-induced injection has been realized by mixed gases of $N_2$ [11]-[12] or $CO_2$ [7] with $He$ to produce electron bunches with acceleration close to the maximum dephasing length to give electron energies up to 1.45 $GeV$, with enhanced yield and small divergence as compared to electron beams from self-injection [7]. Experiments and simulations have demonstrated that ionization-induced injection has a lower intensity threshold $a_0 \sim 2$, (where $a_0 = (eE) / (m\omega c)$ with $e$ the electron charge, $E$ the electric field, $m$ the electron mass, $\omega$ the angular frequency, and $c$ the light speed) as compared with self-injection which requires higher laser intensity of the order of $a_0 \sim 4$ [13]. The addition of higher Z gases is capable of generating electron beams with high charge and low emittance. Recent experiments have reported quasi-monoenergetic electron beams with maximum energy above $0.5\ GeV$ and $2\ mrad$ divergence using 1-cm-long supersonic gas jet of pure nitrogen at electron densities of $\sim 2.2 \times 10^{-18}\ cm^{-3}$ to $1.1 \times 10^{19} cm^{-3}$ [14].

In this paper, we report on electron beam acceleration in pure nitrogen gas jet at relatively low plasma densities of $0.8 \times 10^{18}\ cm^{-3}$ to $2.7 \times 10^{18}\ cm^{-3}$, leading to electron bunches with energies typically 130 MeV. We used a relatively large laser focal spot size ($\sim 27\ \mu m$) and a modest laser intensity ($\sim 3.5 \times 10^{18}\ W \cdot cm^{-2}$), which led to stable laser propagation and efficient ionization-induced injection. The influence of the laser parameters on the acceleration structures (laser-plasma channels), electron beam energy, yield and divergence angle are all well-characterized.

## II. DESCRIPTION OF LWFA EXPERIMENT

The experimental set-up used our electron acceleration is shown schematically in Fig. 1. The experiment was performed using a CPA (chirped pulse amplification)-based Ti: sapphire laser system at Key Laboratory of Laser Plasmas, Shanghai Jiao Tong University, China. Currently, the laser system delivers laser pulses of $30\ fs$ in duration with a peak power of about $50\ TW$. The laser beam was focused in vacuum by using $f/20$ off-axis parabolic mirror (TYDEX Research and Industrial Optics), the focal point was placed above the front edge of 4-mm long slit-shaped gas jet nozzle. The nozzle generates a supersonic gas flow with a Mach number of 5 (Smartshell Co., Ltd A6 series). The gas density was controlled by varying the gas-jet stagnation pressure [15]. The laser-plasma interaction was observed through a top-view imaging of the scattering laser light. A band-pass filter was placed in front of the imaging 14-bit CCD camera to allow only the second-harmonic scattered radiation from the plasma to pass through. After the gas jet, the accelerated electron beam propagates through an aluminum foil, which blocks the transmitted laser light from reaching the detection system. Then an integrating current transformer was installed to measure the electron beam charge. We used a $0.95\ Tesla$, $6\ cm$-long permanent dipole magnet and a fluorescent DRZ screen to capture the electron beam spatial profile and energy spectrum. The magnetic field disperses the electrons in the horizontal plane only, while the vertical beam size gives the beam divergence. A circular hole (diameter $\sim 5\ mm$) made in a $2\ mm$-thick aluminum plate was placed at the entrance of the magnet to serve as a collimator. The electron beam energy spectrum on the DRZ was captured using a synchronized 14-bit charge-coupled-device (CCD) camera.

## III. EXPERIMENTAL RESULTS AND DISCUSSIONS

The laser focal spot size in vacuum was measured to be $\sim 27\ \mu m$ in the full width at half maximum as shown in Fig. 2. The focal spot size was measured (at low laser power) by a removable diagnostic system consists of a microscope lens and a laser beam profiler. Our experiment was performed by using 30 $fs$, 50 $TW$ laser pulses at the wavelength of 800 $nm$. The maximum laser intensity and the corresponding normalized vector potential $a_0$ were $\sim 3.5 \times 10^{18}\ W \cdot cm^{-2}$ and 1.3, respectively. The spatial laser pulse length $L$ corresponds to 9 $\mu m$ and the Rayleigh length is 3.5 $mm$. For the plasma density range from $0.8 \times 10^{18}\ cm^{-3}$ to $5.4 \times 10^{18}\ cm^{-3}$, the plasma wavelength $\lambda_p$ range is from 14.4 to 37.3 $\mu m$. Our experiment is in the linear LWFA scheme, where the spatial laser-plasma parameters satisfy $r_0 \sim 2 - 3L \sim \lambda_p$. The plasma density is low, so the dephasing length is long; the dephasing length defined by $L_d = \lambda_p^3/\lambda^2$, is longer than the gas jet length (4 $mm$). That means the electron beam energy doesn't drop due to dephasing. Fig. 3(a) shows a single, long interaction plasma-channel extending to almost one $Z_R$ achieved through the relativistic self-focusing of the laser pulse. This channel has a diameter around 45 $\mu m$ (this is an upper limit set by the resolution of the top-view imaging system) and length $\sim$ 4 $mm$. They were observed in almost all the laser shots, within nitrogen gas density range 0.5 to 3 $MPa$. Fig. 3(b) shows the spatial profile of a well-collimated electron beam at the plasma density of $\sim 1.1 \times 10^{18}\ cm^{-3}$. The blue (red) curve is the horizontal (vertical) lineout of the electron beam profile. The horizontal and vertical divergences were 3 $mrad$ and 2.8 $mrad$ (FWHM), respectively.

If $P_L \gg P_C$, a long interaction channel or filamentary channels extending to several $Z_R$ through RSF was expected. When the plasma density was above $2.7 \times 10^{18}\ cm^{-3}$, filamentation instability, which may influence the acceleration process and leads to a degraded-quality electron beams, tends to be enhanced. Researches show that it's very important to optimize the laser parameters, such as laser pulse duration and the focal spot. Otherwise, it results in laser beam filamentation and large-divergence electron beams [16], as shown in Fig. 3(d). The drive laser pulse will lose all of its energy after propagating a depletion length. At higher densities, the depletion length is shorter. We measured the plasma channel length at different plasma densities and laser energies; the plasma channel is longer at low plasma density and high laser energy, which is shown in Fig. 3(c). Therefore, the laser-plasma interaction length was longer and it's more likely to accelerate electron beam to high energy. In the electron acceleration experiments based on LWFA, the background plasma electrons can be self-trapped in the plasma wave and accelerated by it. After the laser pulse is intense enough to propagate longer through relativistic self-focusing (RSF), a strong wakefiled is produced at densities high-enough for self-trapping. Fig. 4(a) shows typical quasimonoenergetic electron energy spectra at different laser power. These shots are obtained at the same plasma density of $1.1 \times 10^{18}\ cm^{-3}$. The electron beam is deflected in the horizontal direction while the vertical profile manifests the transverse divergence of the electron beams. The mono-energetic peak energy in the 4$^{th}$ panel is $130\ MeV$. And the quasimonoenergetic and well-collimated electron beam is observed in 45% of the total shots. Also, the long lower energy tails (in the 2$^{nd}$ and 4$^{th}$ panel of Fig. 4) behind the high energy peaks demonstrate the continuous injection and the acceleration of the K-shell electrons of the nitrogen gas. When the laser intensity is high (Fig. 4 the 4$^{th}$ panel), the injection and acceleration of K-shell electrons of the nitrogen is continuous and it leads to a broad electron energy spectrum. Lowering the laser intensity can limit the distance over which the laser pulse is intense enough to ionize the K-shell electrons into the wakefield. When the injection is ceased, the wakefield will continue to accelerate the previously-trapped electrons. If enough amount of acceleration happens after the injection of electrons has ceased, then there will be a low energy tail in the energy spectrum [12], as shown in Figure 4. Our experimental results can be explained by a combination of the self-injection and the ionization induced injection. Since the plasma density is low, the dephasing length and the acceleration length are extended. An optimally large focal spot size can lead to stable laser propagation and electron beam generation [17]. In previous experiments with $TW$ laser pulses, the threshold density for electron self-injection was $n_e \sim 3 \times 10^{18}\ cm^{-3}$ [18]. At lower density, fully blown out bubbles of larger radius are required. There are two methods for reducing the threshold density. One is referred as the ionization-induced injection. Stable self-guiding is realized when the laser spot size is matched with the bubble radius, $R_b$, defined by the relation: $w_0 \sim R_b = \frac{2\sqrt{a_0}}{k_p} \propto n_e^{-1/2}$, where $w_0$ is the Gaussian laser spot radius and $k_p$ is the propagation constant of the plasma wave. As the plasma density decreases, the blown out bubble increases in size and eventually exceeds the laser spot size at low density, which indicates that the self-guiding is unstable. The K-shell and L-shell electrons, requiring $I_L \sim 10^{19}\ W/cm^{-2}$ and $I_L \sim 2 \times 10^{16}\ W/cm^{-2}$ to be ionized, they have a large difference in ionization potential. Therefore, the L-shell electrons are pre-ionized by the leading front of the laser pulse and form the electron sheath of the bubble, while the K-shell electrons are ionized by the peak of the laser pulse. If the bubble radius is very small, more electrons are expelled transversally. Then those electrons circulate around to the inside back edge of the bubble. The electron trapping and acceleration are achieved inside the bubble. On the other hand, if the bubble size is large, more electrons go through the front edge of the cavity. Then the electrons can be injected and accelerated to high energies. In this case, the injection is continuous, indicating the long tail feature of the electron energy spectrum. If the plasma density is too low, the large bubble size and the focal spot size result in laser intensity that is not enough to ionize the K-shell electrons, thus turning off the injection.

Fig. 5 shows the dependence of electron beam divergence (transverse to laser polarization), beam charge and the beam mono-peak energy on laser energy. As the laser energy increases, the beam charge, beam divergence and the mono-peak energy tend to increase. The possible explanation is that as laser intensity increases, electrons are ionized further off the axis and suffer a larger radial ponderomotive force from the laser which causes larger beam divergence.

## IV. CONCLUSIONS

In conclusion, the experiments have manifested that a quasimonoenergetic electron beams with peak energy $\sim 130\ MeV$ and low divergence $\sim 3\ mrad$ can be realized by a $4-mm-long$ pure nitrogen gas jet at low plasma density. Optimization of the laser parameters, such as laser focal spot size and laser intensity, lead to stable laser propagation and the acceleration of high-quality electron beams. Also, the progress of ionization-induced injection can be affected, which can cause either continuous energy spectra or quasimonoenergetic spectra with long-tail feature. We anticipate that laser-driven pure nitrogen gas jets can generate high-quality electron beams through the ionization induced injection scheme of laser wakefiled acceleration at higher laser intensities and lower plasma densities.

**Figure Captions**

Figure 1. Experimental set-up

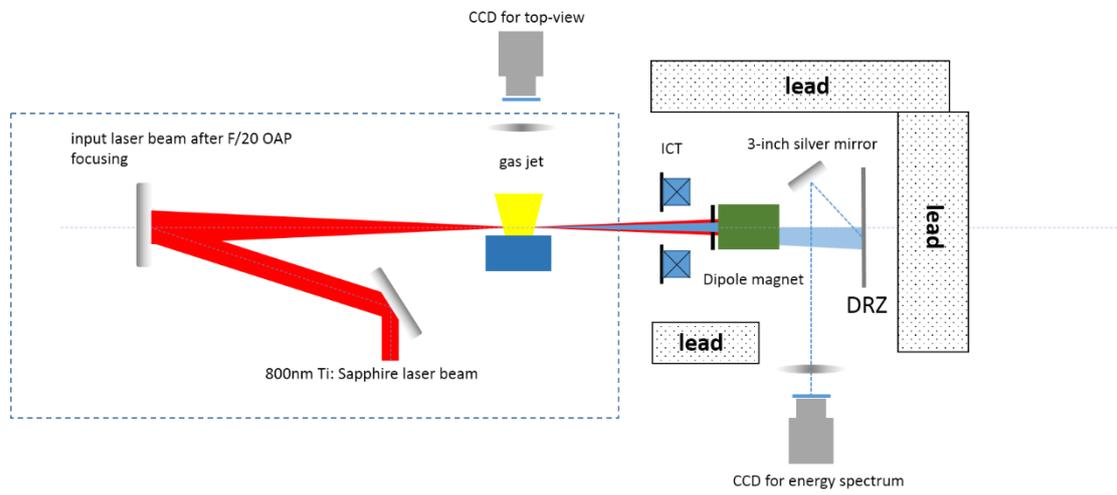

Figure 2. The transverse spatial profiles of a focused power-attenuated laser pulse in vacuum above. The blue (red) curve is the horizontal (vertical) lineout. The horizontal and vertical spot sizes at the entrance were 27 $\mu m$ and 28 $\mu m$.

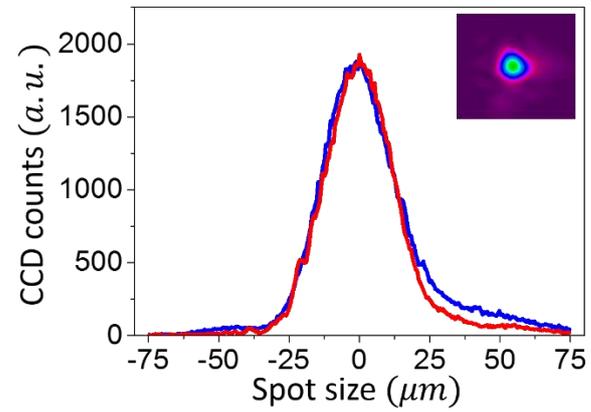

Figure 3. (a) Top-view image of the laser-plasma channel. (b) The transverse spatial profiles of a collimated electron beam generated at the plasma density of $1.1 \times 10^{18} \ cm^{-3}$ (without the permanent dipole magnet). (c) The dependence of the plasma channel length on laser energy at different plasma density. (d) Double-channel phenomenon and corresponding filamentary beams. Channel 1 is shorter than channel 2, which indicates channel 1 causes the electron beam 1 with a divergence of $.52 \ mrad$. The electron beam 2 (divergence ~ $6.23 \ mrad$) has a better quality.

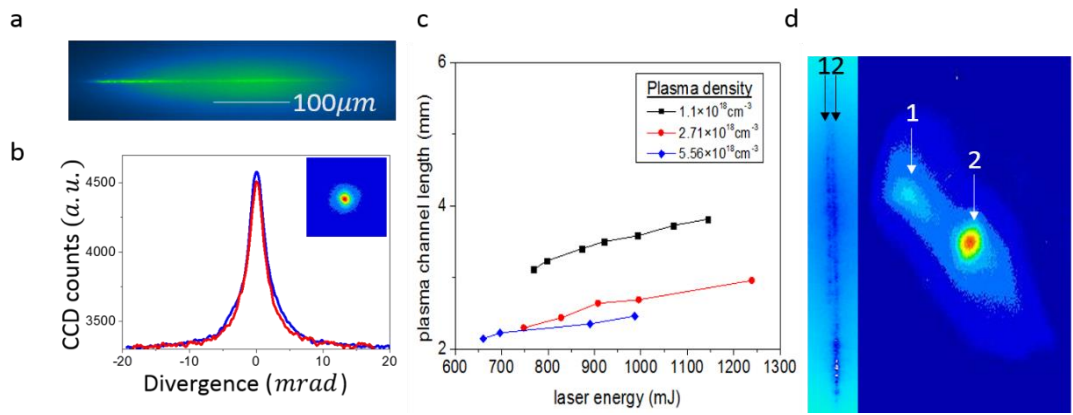

Figure 4. The electron beam energy spectra for different laser energies. The laser energies (top-bottom panels) are 0.79 J, 0.90 J, 1.05 J and 1.38 J respectively.

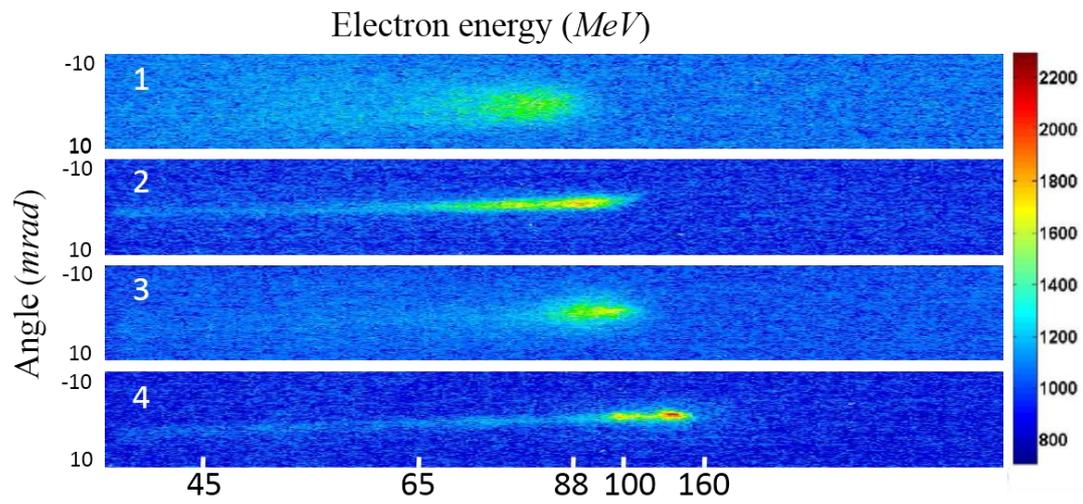

Figure 5. The dependence of the electron beam divergence angle, yield (charge), mono-peak energy on the laser energy is shown in (a), (b) and (c), respectively.

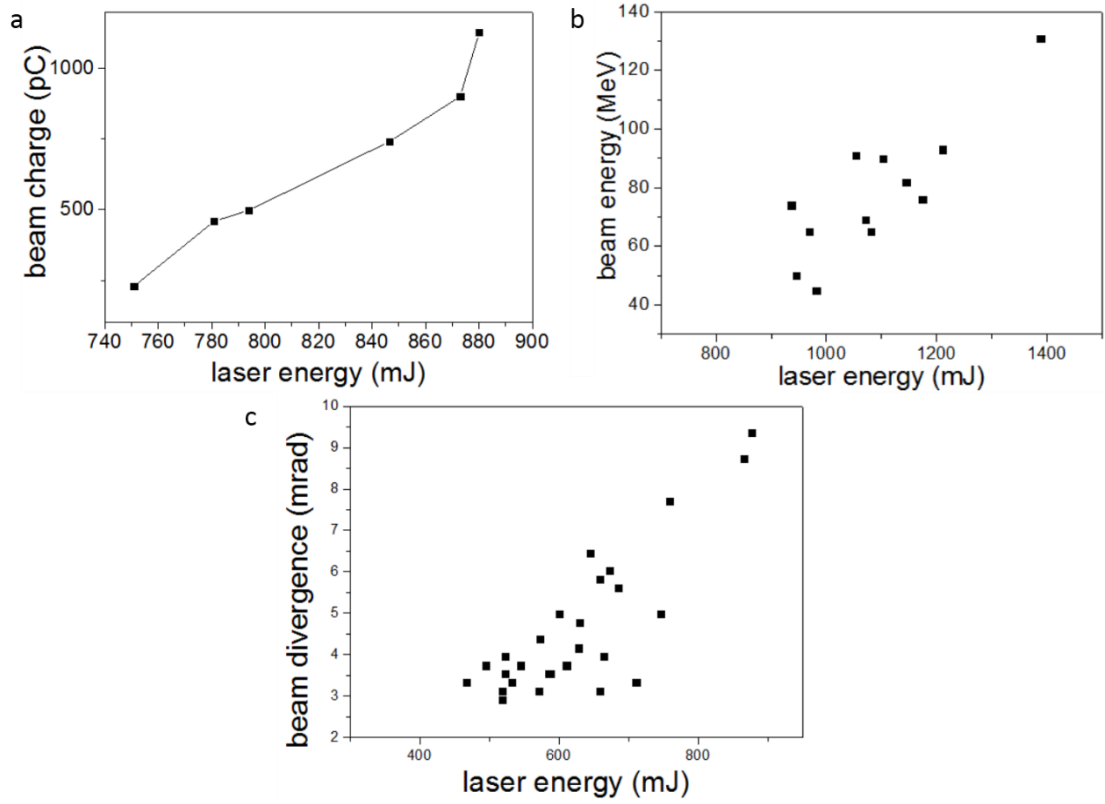